\documentclass[prd,twocolumn,showpacs,superscriptaddress,preprintnumbers,nofootinbib, amsmath,amssymb]{revtex4-1}

\usepackage{tikz}
\usetikzlibrary{arrows}

\usepackage[colorlinks=true,urlcolor=blue,linkcolor=blue,citecolor=blue]{hyperref}
\usepackage[normalem]{ulem}
\usepackage[capitalize]{cleveref}
 
\usepackage{comment}
\usepackage{flushend}

\usepackage[T1]{fontenc} 

\usepackage{mathtools}
\usepackage{xcolor}
\usepackage{graphicx}
\usepackage{dcolumn}
\usepackage{bm}
\usepackage{multirow}
\usepackage{physics}

\usepackage{epstopdf}

\newcommand{\Beq}{\begin{equation}\begin{aligned}}
\newcommand{\Eeq}{\end{aligned}\end{equation}}

\begin{document}

\title{Dark Bondi Accretion Aided by Baryons and the Origin of JWST Little Red Dots}

\author{Wei-Xiang Feng}
\email{wxfeng@mail.tsinghua.edu.cn}
\affiliation{Department of Physics, Tsinghua University, Beijing 100084, China}

\author{Hai-Bo Yu}
\email{haiboyu@ucr.edu}
\affiliation{Department of Physics and Astronomy, University of California, Riverside, CA 92521, USA}

\author{Yi-Ming Zhong}
\email{yiming.zhong@cityu.edu.hk}
\affiliation{Department of Physics, City University of Hong Kong, Kowloon, Hong Kong SAR, China}


\begin{abstract}
The gravothermal core collapse of self-interacting dark matter halos provides a mechanism for seeding supermassive black holes in the early Universe. We show that the collapse of a small fraction of the halo mass can trigger general-relativistic instability and produce black hole seeds in halos with masses $\gtrsim10^{9}{\rm\,M_\odot}$ at high redshifts. We investigate whether this process can account for the origin of JWST little red dots at $z\sim4\textup{--}11$, which host black holes with masses $\sim10^7{\rm\,M_\odot}$. We find that such masses may be reached within $\sim500{\rm\,Myr}$ through rapid core collapse followed by efficient accretion, even for initial seed masses less than $10^3{\rm\,M_\odot}$. This scenario allows a substantial fraction of the black hole mass to originate from dark matter accretion, potentially offering a viable pathway to forming massive black holes at high redshifts.
\end{abstract}

\maketitle

{\bf Introduction.} The origin of supermassive black holes (SMBHs) remains a mystery, particularly those found at high redshifts and associated with luminous quasars~\cite{Mortlock:2011gaa,Wu:2015ula,Banados:2017unc,Matsuoka:2018iiz,Wang:2018puv,Yang:2020uga,Bogdan:2023ilu}. It is puzzling how a black hole could grow to a mass of $10^9{\rm\,M}_\odot$ when the Universe was only about $5\%$ of its present age. In~\cite{Feng:2020kxv}, we proposed a scenario to explain the origin of SMBHs associated with high-luminosity quasars, through the gravothermal collapse of self-interacting dark matter (SIDM) halos~\cite{Balberg:2001qg,Balberg:2002ue}. As a self-gravitating system, an SIDM halo has a negative heat capacity, and its central region, corresponding to $(0.1\textup{--}1)\%$ of the total mass, can collapse into an ultradense core, which can then further evolve into a black hole. We showed that the presence of compact central baryons can accelerate the onset of collapse and the self-interactions can dissipate the remnant angular momentum  of the collapsed region. The collapse of massive halos of $10^{11}{\rm\,M}_\odot$ at redshifts $z\gtrsim10$ could produce a black hole with a mass of $10^8\textup{--}10^9{\rm\,M}_\odot$; see also~\cite{Pollack:2014rja,Choquette:2018lvq,Xiao:2021ftk,Meshveliani:2022rih,Roberts:2024wup}. 

More recently, observations from the James Webb Space Telescope (JWST) have revealed compact active galactic nuclei at redshifts $z\sim4\textup{--}11$, hosting SMBHs with masses of $10^6\textup{--}10^8{\rm\,M}_\odot$~\cite{Kocevski:2023aa,Harikane:2023aa,Matthee:2023utn}. These systems, referred to as little red dots (LRDs), exhibit unusually high black hole-to-stellar mass ratios of $\gtrsim10^{-3}\textup{--}10^{-1}$\,\cite{Pacucci:2023oci}, significantly larger than those observed in the local Universe~\cite{Kormendy:2013dxa}. This suggests that black holes in LRDs are overmassive relative to their host galaxies and likely originate from heavy seeds (e.g.,~\cite{Natarajan:2023rxq}). Nevertheless, the detailed formation pathways of LRDs remain uncertain.

\begin{figure}[t]
\centering
   \includegraphics[width=0.3\textwidth]{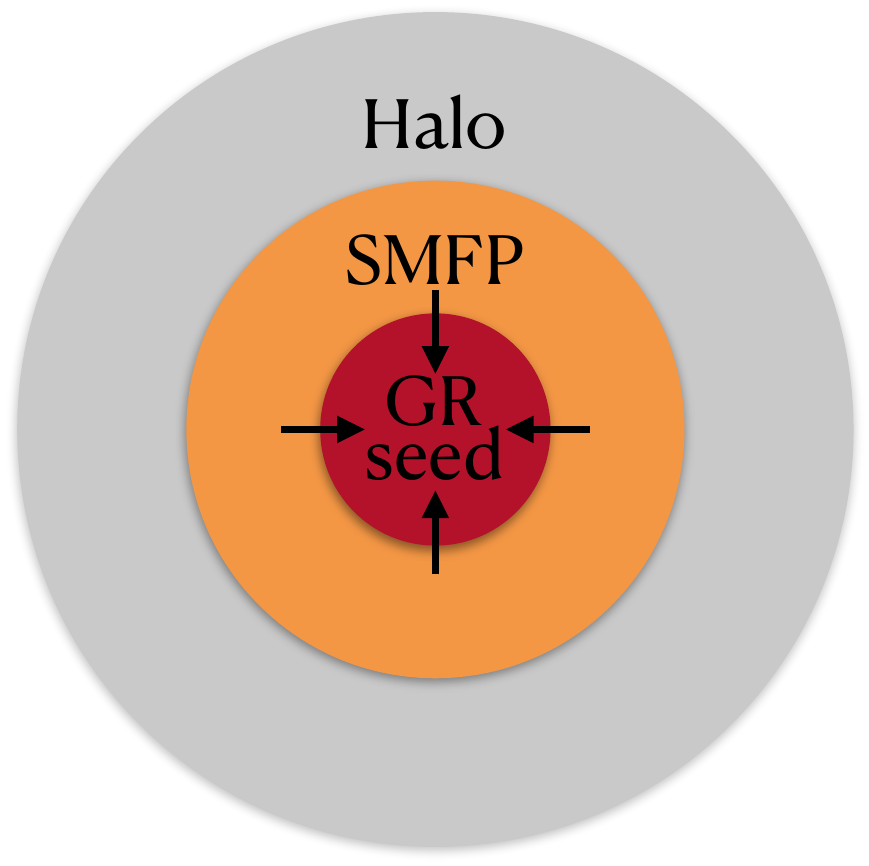}
    \caption{A schematic illustration of dark matter accretion. The central region of a collapsed self-interacting dark matter halo forms a seed black hole through general-relativistic (GR) instability, which subsequently grows by accreting dark matter particles from the short-mean-free-path (SMFP) region. Not to scale.}
   \label{fig:scheme}
\end{figure}

In this work, we explore the SIDM-based seeding mechanism for the formation of LRDs and show that they may arise from the gravothermal collapse of $\gtrsim10^9{\rm\,M}_\odot$ halos. We demonstrate that the collapse of SIDM halos is robust to baryonic feedback associated with bursty galaxy formation. We also address the crucial issue of how the entire collapsed central halo can evolve into a black hole, which was an assumption in earlier studies. To trigger general-relativistic instability, the 1D central velocity dispersion of dark matter particles must reach approximately $0.3\,c$~\cite{Balberg:2001qg,Feng:2021rst}. Only a tiny fraction of the halo mass, on the order of $10^{-10}\textup{--}10^{-8}$, can attain this condition following the trajectory of gravothermal evolution~\cite{Balberg:2001qg,Gad-Nasr:2023gvf}. For a $10^9{\rm\,M}_\odot$ halo, the initial seed black hole mass is expected to be on the order of $1{\rm\,M_\odot}$. 

We show that such a solar-mass seed may grow to $10^7{\rm\,M_\odot}$ within $500{\rm\,Myr}$ through accretion. The initial growth phase is dominated by baryonic Eddington accretion, during which the black hole increases in mass from $1{\rm\,M_\odot}$ to $10^4{\rm\,M_\odot}$. After this stage, dark Bondi accretion becomes the dominant channel, efficiently consuming SIDM particles from the ultradense core and driving the black hole mass up to $10^7{\rm\,M_\odot}$. In this scenario, the black hole may grow primarily through dark matter accretion rather than baryonic processes. This dark Bondi accretion mechanism could also operate for $1\textup{--}100\,{\rm M_\odot}$ seeds produced by Population III stars that populate the high-redshift Universe~\cite{Regan:2024wsu}. We further demonstrate that Bondi accretion in the dark sector is not limited by a dark Eddington bound for viable SIDM particle models. Fig.~\ref{fig:scheme} illustrates the process of dark matter accretion.

{\bf Gravothermal collapse timescale.} We consider a Navarro--Frenk--White (NFW) halo formed at redshift $z$, and its density profile is given by $\rho_{\rm NFW}(r)=\rho_s(r/r_s)^{-1}\left(1+r/r_s\right)^{-2}$~\cite{Navarro:1996gj}, where the scale density $\rho_s$ and radius $r_s$ are
\begin{align} 
\rho_s = \frac{200\,\rho_c\,c_{200}^3}{3\,f(c_{200})}, 
\ r_s = \left(\frac{3\,M_{200}}{800\pi\,c_{200}^3\,\rho_c}\right)^{1/3},
\label{eq:rhosrs}
\end{align} 
respectively. $M_{200}$ is the virial mass within the radius $r_{200}$, $c_{200}\equiv r_{200}/r_s$ is the halo concentration, $f(c_{200})\equiv\ln(1+c_{200})-c_{200}/(1+c_{200})$, and the critical density
\begin{equation} 
\rho_c(z) = \frac{3H_0^2}{8\pi G}\left[\Omega_{m,0}(1+z)^3 + \Omega_{\Lambda}\right].
\label{eq:rhoc}
\end{equation}
We take $H_0 = 67.6\,{\rm km/s/Mpc}$, $\Omega_{m,0} = 0.312$, and $\Omega_{\Lambda}=0.688$ from Planck cosmology~\cite{Planck:2018vyg}. We can write $M_{200}=M_0f(c_{200})$, where $M_0$ is the fiducial mass of the NFW halo $M_0\equiv4\pi\rho_sr_s^3$. 
 
Assuming that dark matter self-interactions set to operate for the halo formed at redshift $z$, the core-collapse timescale can be estimated as~\cite{Essig:2018pzq}
\begin{align} 
&t_{\rm cc}(z)=\frac{150}{C}\frac{t_0}{\rho_s r_s\left(\sigma/m\right)} \nonumber\\
\approx&495{\rm\,Myr}
\left(\frac{f(c_{200})}{f(3)}\right)^{3/2}\!\!\left(\frac{c_{200}}{3}\right)^{-7/2}\left(\frac{M_{200}}{10^{9}{\rm\,M}_\odot}\right)^{-1/3} \nonumber\\
&\times\left[\frac{\Omega_{m,0}\left(1+z\right)^3+\Omega_\Lambda}{\Omega_{m,0}\left(1+10\right)^3+\Omega_\Lambda}\right]^{-7/6}\!\!\left(\frac{\sigma/m}{150{\rm\,cm^2/g}}\right)^{-1},
\label{eq:collapsetime}
\end{align}
where $C$ is a numerical coefficient and we take $C\simeq0.75$ as calibrated from $N$-body simulations~\cite{Zhong:2023yzk}, $t_0\equiv(4\pi G\rho_s)^{-1/2}$ is the fiducial timescale, and $\sigma/m$ is the dark matter self-interacting cross section per unit particle mass, which should be interpreted as an effective cross section for the given halo mass in the context of velocity-dependent SIDM models~\cite{Yang:2022hkm,Yang:2022mxl}.

The halo concentration becomes insensitive to the halo mass at redshifts $z\gtrsim5$ and its median value asymptotes to $c_{200}\approx3$~\cite{Ishiyama:2020vao}. As an estimate for LRDs, we consider a halo with a median concentration of $c_{200} = 3$ and a mass of $M_{200}\approx10^{9}{\rm\,M}_\odot$ formed by $z=10$, i.e., $t=470{\rm\,Myr}$ after the Big Bang. The required cross section is $\sigma/m\sim150{\rm\,cm^2/g}$ and $t_{\rm cc}\sim500{\rm\,Myr}$. A black hole with mass $10^{-2}M_{200}\sim10^7{\rm\,M_\odot}$ could form at $t\sim970{\rm\,Myr}$ ($z\sim6$), assuming that the entire short-mean-free-path (SMFP) region evolves into the hole. If the halo has a relatively higher concentration, the cross section could be lower, e.g., $\sigma/m\approx50{\rm\,cm^2/g}$ for $c_{200}\approx5$. Indeed, more recent simulations show that the presence of a high-concentration tail is robust for halos at high redshifts, and the concentration can reach $\approx4\textup{--}10$ for $z\sim6\textup{--}10$~\cite{Yung:2023sib}. Taking this into account, Ref.~\cite{Jiang:2025jtr} investigates the formation of LRDs in the SIDM framework and obtains constraints on SIDM models. In our numerical calculations presented below, we adopt $c_{200}=3$ as a fiducial value, which should be regarded as a conservative choice.

Importantly, the presence of baryons modifies the collapse timescale. On one hand, baryons can deepen the gravitational potential and accelerate the collapse accordingly~\cite{Elbert:2016dbb,Feng:2020kxv,Sameie:2021ang,Zhong:2023yzk,Despali:2025koj}. The acceleration becomes stronger as the baryonic distribution becomes more compact. Since LRD galaxies are surprisingly compact~\cite{Maiolino:2023bpi,Baggen:2024vef}, this effect can be significant. On the other hand, baryonic feedback associated with bursty star formation may inject energy into the central halo, which could delay the collapse or even prevent it. Because star formation is particularly active in early galaxies, it is crucial to demonstrate that SIDM core collapse is robust to baryonic feedback. In this work, we investigate both competing effects.

We use the conducting fluid model with a static baryonic potential as in~\cite{Feng:2020kxv,Zhong:2023yzk}, and extend it to include an oscillating external potential that characterizes the bursty feedback effect as motivated by~\cite{Pontzen:2011ty}. To be concrete, we fix model parameters by consider the LRD~ID\,61888 with a black hole mass of $1.66\times10^7{\rm\,M}_\odot$ at $z\approx5.87$~\cite{Maiolino:2023bpi}. Its stellar surface density is characterized by a S\'ersic profile with index $n=0.9$, half-light radius $R_\star=0.09{\rm\,kpc}$, and stellar mass $M_\star=1.29\times10^8{\rm\,M}_\odot$. We fit the stellar distribution with the Plummer profile $\rho_B(r)=(3M_B/4\pi a^3)\left(1+r^2/a^2\right)^{-5/2}$ with scale radius $a=0.1{\rm\,kpc}$ and total mass $M_B=1.4\times10^{8}{\rm\,M}_\odot$ to model the baryonic potential. We assume a halo at $z=10$ with $M_{200}=1.5\times10^9{\rm\,M}_\odot$ and $c_{200}=3$, corresponding to $\rho_s\approx1.5\times10^8{\rm\,M}_\odot/{\rm kpc^3}$ and $r_s\approx1.1{\rm\,kpc}$. The fiducial time is $t_0=1/\sqrt{4\pi G\rho_s}\approx10.9{\rm\,Myr}$ and the fiducial mass $M_0=4\pi\rho_sr_s^3\approx2.4\times10^9{\rm\,M_\odot}$. Hence, $M_B/M_0=0.06$ and $a/r_s=0.1$ for the LRD~ID\,61888 system. For comparison, we also consider a case where we reduce the size by half $a/r_s=0.05$, while keeping $M_B$ fixed. Our simulations take two values for the fiducial cross section $\hat{\sigma}\equiv(\sigma/m)\rho_s r_s=0.5$ and $1.0$, corresponding to $\sigma/m= 15{\rm\,cm^2/g}$ and $30{\rm\,cm^2/g}$, respectively.  

Besides the static potential matched to the observed stellar distribution of the LRD~ID\,61888, we also introduce an oscillating baryonic potential to mimic repeated cycles of feedback-driven blowout and subsequent gravitational re-accretion.
\begin{equation}
\rho_B (r,t) = \frac{3M_B}{8\pi a^3\left(1+r^2/a^2\right)^{5/2}}\left[1+\cos \left( \frac{2\pi t}{P}\right)\right]
\end{equation}
where we fix $M_{B}$ and $a$ to be the same as in the static potential; $P$ is the oscillation period, which we set equal to the dynamical time of the halo, $P=t_0$. This is a reasonable assumption, reflecting that the timescale for baryonic cycles is set by the halo gravity.

\begin{figure}[t]
\centering
   \includegraphics[width=0.48\textwidth]{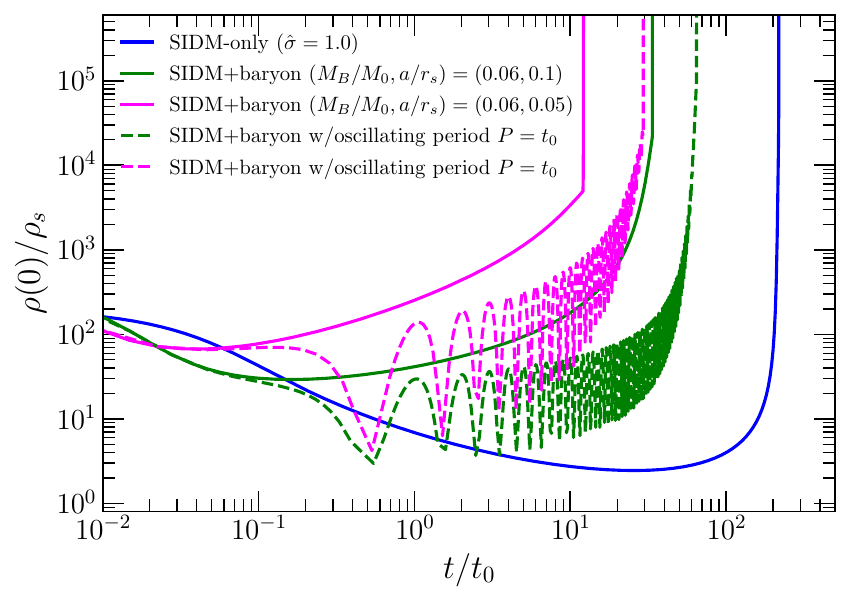}
    \caption{Evolution of the normalized central density for simulated halos without a baryonic potential (blue), and with static (solid) and oscillating (dashed) Plummer baryonic potentials, assuming $(M_B/M_0, a/r_s) = (0.06, 0.1)$ (green) and $(0.06, 0.05)$ (magenta). The fiducial self-interaction cross section is $\hat{\sigma} = 1.0$. The presence of a baryonic potential enhances the central density and accelerates gravothermal evolution, with a more compact baryonic distribution producing a stronger effect, while SIDM core collapse remains robust to baryonic feedback.}
   \label{fig:coll_time}
\end{figure}

Fig.~\ref{fig:coll_time} shows the evolution of the central density $\rho (0)/\rho_s$ of the simulated halos for $\hat{\sigma} = 1.0$. For the static potential, the collapse time is reduced by factors of $7$ and $20$ for $a/r_s=0.1$ (solid green) and $0.05$ (solid magenta), respectively, compared to the SIDM-only simulation (solid blue). Intriguingly, even with an oscillating baryonic potential, the core collapse does not stall but still proceeds faster than in the SIDM-only case, with reduction factors of $3$ for $a/r_s=0.1$ (dashed green) and $7$ for $0.05$ (dashed magenta) due to energy injection. The central halo density oscillates in response to the oscillating potential, but the amplitude diminishes at late times as the central density builds up during collapse and the collapsed region remains well localized near the halo center. These fluid simulations demonstrate that SIDM core collapse is robust to baryonic feedback. We have verified that as the halo concentration increases, the impact of baryonic feedback on the collapse time decreases relative to the results shown in Fig.~\ref{fig:coll_time}, which assume a median-concentration halo. Since high-concentration halos are naturally favored to produce black hole seeds in the SIDM-seeding scenario, their collapse should be even more robust against baryonic feedback as demonstrated in recent $N$-body simulations~\cite{Kong:2026piq}.

Our calculation can be improved in several respects. The feedback model assumes oscillations of the entire stellar disk throughout the SIDM gravothermal evolution, which likely overestimates the feedback effect, as the bursty phase of star formation should weaken once most of the gas has been converted into stars. High-redshift galaxies may exhibit disky stellar structures~\cite{Kuhn:2024}, deviating from the spherical symmetry assumption. In addition, we do not include the gas contribution, which could have a steep density profile due to dissipation. We also neglect the backreaction of SIDM core collapse on the baryonic distribution. Hydrodynamical simulations from the FIRE-2 project show that the stellar distribution becomes more compact in SIDM runs than in CDM runs at $z\approx0$ due to the backreaction~\cite{Sameie:2021ang}. In this regard, the acceleration of core collapse could be even stronger than what we have shown.

While we adopt a fiducial halo mass of $\sim10^9\,{\rm M_\odot}$ an example to highlight the dynamics, it is important to note that the SIDM-seeding scenario applies across the halo mass spectrum. More massive halos may directly form seeds for LRDs, while less massive halos may form seeds that later merge. Quantitative predictions for the LRD population in the SIDM-seeding scenario in the dark matter-only case have been explored in~\cite{Jiang:2025jtr}. Incorporating baryonic physics and its correlation with the angular momentum distributions of baryons and dark matter halos (e.g.,~\cite{Pacucci:2025tyl}), as well as mergers, will be an important next step, which we leave for future work. In what follows, we focus on case studies that elucidate other key physical ingredients of the SIDM-seeding scenario.

{\bf The initial black hole.} In the estimate above, we have assumed that the black hole mass is set by the mass in the SMFP region $M_{\rm smfp}$, which is characterized by the condition of Knudsen number, $Kn\equiv\lambda/H<1$, with $\lambda$ the SIDM mean-free-path and $H$ the local Jeans length. From our fluid simulations, we find $M_{\rm smfp}\sim10^{-2} M_0$; see~\ref{sec:app1} and also~\cite{Feng:2020kxv}. However, not all the mass in the SMFP region could collapse into a black hole instantaneously. In fact, the general-relativistic instability can only be triggered in the very central region, where the 1D velocity dispersion of SIDM particles reaches $\nu_{\rm GR}\approx0.3\, c$~\cite{Balberg:2001qg,Feng:2021rst}. The mass of this central region, denoted as $M_{\rm GR}$, can be estimated using the scaling relation $M_{\rm GR}\approx(\nu/\nu_{\rm GR})^{2\alpha}M_{\rm smfp}$ with $0.85\lesssim\alpha\lesssim1$, which is derived from the trajectory of gravothermal evolution~\cite{Balberg:2001qg,Gad-Nasr:2023gvf}. For halos with $M_{200}\sim10^8\textup{--}10^{11}{\rm\,M_\odot}$, $M_{\rm GR}/M_{200}\approx10^{-10}\textup{--}10^{-8}$~\cite{Gad-Nasr:2023gvf}. For LRDs, $M_{200}\sim10^{9}{\rm\,M_\odot}$ and $M_{\rm GR}\sim 1{\rm\,M}_\odot$. Note that $M_{200} = f(c_{200}) M_0 \sim M_0$, since $f(c_{200})$ is typically an ${\cal O}(1)$ factor.

The scaling relation relies on the extrapolation to the relativistic regime from fluid simulations. Here, we provide a complementary argument from the perspective of energy conservation~\cite{Ralegankar:2024zjd}. Assuming the SMFP region itself is virialized, we have $K=|E|=-U/2$, where $K$ is the kinetic energy, $|E|$ is the binding energy and $U$ is the potential energy. $U=-0.6GM_{\rm smfp}^2/r_{\rm smfp}$, assuming a uniform density of the SMFP region, and hence $K=|E|=0.3GM_{\rm smfp}^2/r_{\rm smfp}$. When the instability is triggered at the halo center, the associated binding energy is $0.035M_{\rm GR} c^2$~\cite{Feng:2021rst}. Meanwhile, the binding energy of the remaining part of the SMFP region is $0.3GM_{\rm rem}^2/r_{\rm rem}$. From the condition of energy conservation, we have $0.035M_{\rm GR} c^2+0.3GM_{\rm rem}^2/r_{\rm rem}=0.3GM_{\rm smfp}^2/r_{\rm smfp}$, which is equal to the binding energy in magnitude. Therefore, we have $0.035M_{\rm GR} c^2\lesssim 0.3GM_{\rm smfp}^2/r_{\rm smfp}$, and 
\begin{equation}
\frac{M_{\rm GR}}{M_{\rm smfp}}\lesssim
10\times\left(\frac{10^{-2}r_s}{r_{\rm smfp}}\right)\left(\frac{M_{\rm smfp}}{10^{-2}M_0}\right)\frac{GM_0}{c^2r_s}.
\label{eq:seedmass}
\end{equation}
If $M_0\approx10^9{\rm\,M}_\odot$ and $r_s\approx1{\rm\,kpc}$ for LRDs, we obtain $M_{\rm GR}/M_{\rm smfp}\lesssim10^{-7}$, corresponding to $M_{\rm GR}\sim1{\rm\,M}_\odot$. For $M_0\simeq10^{12}{\rm\,M}_\odot$ and $r_s\simeq10{\rm\,kpc}$, we obtain $M_{\rm GR}\sim10^5{\rm\,M_\odot}$. 
These estimates are broadly consistent with those from~\cite{Gad-Nasr:2023gvf}. 

More recently, Ref.~\cite{Gu:2026zzq} incorporates non-equilibrium effects in full GR fluid simulations and shows that the initial black hole mass is much smaller than the mass of the SMFP region, in agreement with our energy conservation argument. Quantitatively, for a $6.3\times10^9\,{\rm M_\odot}$ halo with a concentration close to the median and $\sigma/m=5\,{\rm cm^2/g}$, the initial seed mass is $200\,{\rm M_\odot}$. The seed mass increases with concentration and cross section, while the ratio $M_{\rm GR}/M_{\rm smfp}\approx4\times10^{-6}$ remains nearly constant. This ratio is larger than the estimate in Eq.~\ref{eq:seedmass} by a factor of $10$, likely because non-equilibrium effects are dynamically included in their simulations, indicating that the estimate based on Eq.~\ref{eq:seedmass} is conservative.

We note that the initial seed mass may further increase after accounting for the mass ejected during seed formation, which could be rethermalized in the SMFP region, as well as the presence of an external baryonic potential and the velocity dependence of the cross section—effects that are neglected in the present study. On the other hand, even a $\sim10^2\textup{--}10^3\,{\rm M_\odot}$ seed black hole remains well below $10^6{\rm\,M_\odot}$, the typical threshold for an SMBH, but can still grow substantially via dark matter accretion. In what follows, we discuss dark Bondi accretion, followed by baryonic Eddington accretion.

{\bf Dark Bondi accretion.} Since dark matter in the SMFP region is in the hydrodynamic limit ($Kn<1$), we apply the Bondi accretion ~\cite{Bondi:1944rnk,Bondi:1947fta,Bondi:1952ni} to model the initial seed accreting the surrounding SIDM particles
\begin{equation}
\dot{M}_{\rm Bon}=4\pi\lambda_s\frac{G^2M^2}{a_\infty^3}\rho_\infty,
\label{eq:Bondi_rate}
\end{equation}
where the mass density $\rho_\infty=mn_\infty$ and sound speed $a_\infty$ are evaluated at a distance $r_\infty$ much larger than the Bondi radius $r_{\rm Bon}=2GM/a_\infty^2$. We take $r_\infty$ to be the edge of the SMFP region. The accretion eigenvalue $\lambda_s= \lambda_s(\gamma)$ is an 
$\mathcal{O} (1)$ factor that depends on the adiabatic index $\gamma$ of the accreted SIDM fluid, $\gamma=4/3+\sqrt{1-3\nu_\infty^2/c^2}/3$~\cite{Feng:2022fuk}, where $\nu_\infty$ denotes the 1D velocity dispersion at $r_\infty$ and $\nu_\infty = a_\infty/\sqrt{\gamma} \ll c$. We take $\gamma \approx 5/3$, which gives $\lambda_s \approx 0.25$~\cite{Shapiro:1983du}.

Assuming $\rho_\infty$ and $a_\infty$ are nearly constant during the accretion process, the timescale is obtained by integrating of Bondi accretion rate, i.e.,
\begin{equation}
t_{\rm acc}=\frac{a_\infty^3}{4\pi\lambda_s G^2\rho_\infty}\left(\frac{1}{M_{\rm GR}}-\frac{1}{M_{\rm smfp}}\right).
\label{eq:bondi2}
\end{equation}
At $r_\infty\sim r_{\rm smfp}\sim 10^{-2}r_s$, the sound speed $a_\infty=\sqrt{\gamma}\,\nu_\infty$. We parametrize  the conditions at $r_\infty$ as $\rho_\infty=A\,\rho_s$ and $\nu_\infty=B\,\nu_0$, where $\nu_0 \equiv r_s \sqrt{4\pi G \rho_s}$ is a fiducial velocity. Thus,  $a_\infty=\sqrt{\gamma}\,B\,\nu_0$. Since $M_{\rm GR}\ll M_{\rm smfp}$, we can neglect the term proportional to $1/M_{\rm smfp}$ in Eq.~\ref{eq:bondi2} and express $t_{\rm acc}$ as
\begin{equation}
t_{\rm acc}=8.6\left(\frac{B^3}{A}\right)\frac{M_0}{M_{\rm GR}}t_0
\end{equation}
where $M_0=4\pi \rho_s r^3_s$ and $t_0=1/\sqrt{4\pi G\rho_s}$.  For $M_0\sim10^9{\rm\,M_\odot}$, $t_0\sim 11{\rm\,Myr}$, $M_0/M_{\rm GR}\sim10^9$, and $B^3/A\sim10^{-5}$ from our fluid simulations; see~\ref{sec:app1}, $t_{\rm acc}\sim10^3{\rm\,Gyr}$, which is too long to be relevant. This remains true even if $M_{\rm GR}$ is ${\cal O}(10)$ times larger, as suggested in~\cite{Gu:2026zzq}. For $M_0\sim10^{12}{\rm\,M}_\odot$, $t_0\sim 15{\rm\,Myr}$, $M_0/M_{\rm GR}\sim10^7$, and $B^3/A\sim10^{-7}$~\cite{Feng:2020kxv}. In this case, $t_{\rm acc}\sim129{\rm\,Myr}$, which is comparable to the collapse timescale.

We note that the Bondi accretion timescale calculated above is likely conservative, as the SIDM SMFP core also evolves during accretion. While the velocity dispersion $\nu_\infty$, parametrized by $B$, is bounded by the speed of light, the SIDM core density $\rho_\infty$, parametrized by $A$, can continue to grow and eventually approach the density of the resulting black hole. This evolution can lead to a value of $B^3/A$ much smaller than $10^{-5}$ for the benchmark $10^9{\rm\,M_\odot}$ halos considered for LRDs. The Bondi accretion timescale could therefore be shorter than our estimate based on the assumption of a quasi-static SMFP core.

In addition, for the Bondi accretion to be valid, the Bondi radius must be smaller than the boundary radius of the SMFP region, i.e., $r_{\rm Bon}< r_\infty$. We find that $r_{\rm Bon}=2 GM/a^2_\infty\lesssim(1.2/B^2)(M_{\rm smfp}/M_0)r_s\approx (1.2\times10^{-2}) (10^{-2}) r_s$ in this scenario, which is smaller than $r_\infty\sim10^{-2}r_s$. Thus, it is reasonable to treat the SIDM accretion from the SMFP region as a Bondi problem.

\begin{figure}[t]
\centering
   \includegraphics[width=0.47\textwidth]{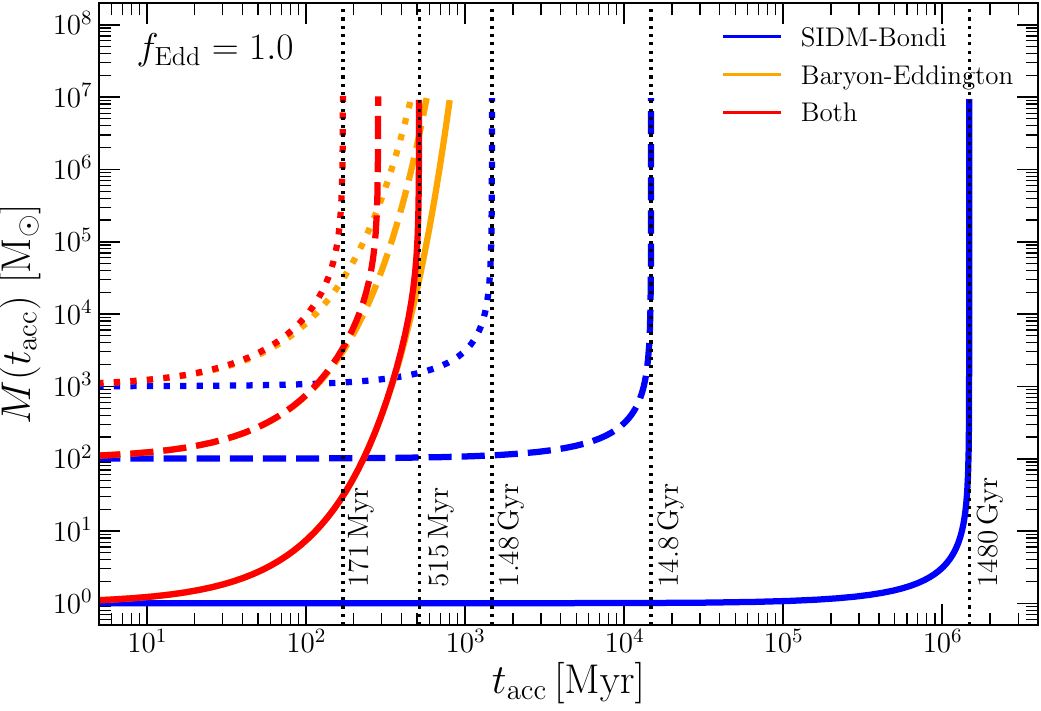}
    \caption{Growth rate of the black hole mass due to dark Bondi accretion (blue), baryonic Eddington accretion (orange), and their combined contribution (red), shown for initial seed masses of $M_{\rm GR} = 1{\rm\,M_\odot}$ (solid), $100{\rm\,M_\odot}$ (dashed), and $1000{\rm\,M_\odot}$ (dotted). Note that the mass of black hole seeds originating from the remnants of Population III stars is $\sim100{\rm\,M_\odot}$.}
   \label{fig:acc_time}
\end{figure}

{\bf Baryonic Eddington accretion.} Intriguingly, observations indicate that baryonic gas is abundant in LRDs~\cite{Naidu:2025rpo}. In this section, we show that baryonic accretion plays a crucial role in enabling the seed black hole to achieve rapid early growth. The rate of Eddington accretion of baryonic gas is 
\begin{equation}
\dot{M}_{\rm Edd}=\left(\frac{1-\epsilon}{\epsilon}\right)f_{\rm Edd}\frac{4\pi GM}{(\sigma_{\rm rad}/m)c},
\label{eq:Edd_rate}
\end{equation}
where $\epsilon$ is the radiative efficiency, $f_{\rm Edd}$ is the accretion efficiency, $\sigma_{\rm rad}$ is the cross section responsible for the radiation pressure. Combining Eqs.~\ref{eq:Bondi_rate} and~\ref{eq:Edd_rate}, we obtain the total accretion rate $\dot{M}=\dot{M}_{\rm Edd}+\dot{M}_{\rm Bon}$, which can be integrated to give the black hole mass as a function of time:
\begin{equation}
M(t_{\rm acc})=\frac{M_{\rm GR}\,e^{t_{\rm acc}/\tau}}{1-\dot{M}_{\rm Bon}/\dot{M}_{\rm Edd}\big|_{M_{\rm GR}}\left(e^{t_{\rm acc}/\tau}-1\right)},
\end{equation}
where $\tau\equiv \left[\left(\frac{1-\epsilon}{\epsilon}\right)f_{\rm Edd}\frac{4\pi G}{(\sigma_{\rm rad}/m)c}\right]^{-1}$ is the Salpeter time~\cite{Salpeter:1964kb}. Taking $\epsilon=0.1$, $\sigma_{\rm rad}/m=0.4{\rm\,cm^2/g}$ for the Thomson cross section, and $f_{\rm Edd}=1$, we have $\tau=50{\rm\,Myr}$ and 
\begin{align}
&\frac{\dot{M}_{\rm Bon}}{\dot{M}_{\rm Edd}}\bigg|_{M_{\rm GR}}=\lambda_s\left(\frac{\epsilon}{1-\epsilon}\right)\frac{\sigma_{\rm rad}/m}{f_{\rm Edd}}\frac{GM_{\rm GR}}{a_\infty^3}\rho_\infty c \nonumber\\
&=\lambda_s\left(\frac{\epsilon}{1-\epsilon}\right)\frac{\sigma_{\rm rad}/m}{f_{\rm Edd}}\left(\frac{A}{B^3}\right)\gamma^{-3/2}\frac{M_{\rm GR}}{M_0}\rho_s ct_0\;.
\label{eq:Bond_to_Edd}
\end{align}
Because the Eddington accretion rate scales linearly with mass $\dot M_{\rm Edd} \propto M$, whereas the Bondi accretion rate scales as $\dot M_{\rm Bon} \propto M^2$, baryonic Eddington accretion dominates in the early growth phase when the seed mass is small as $\dot{M}_{\rm Bon}/\dot{M}_{\rm Edd}\big|_{M_{\rm GR}}\ll1$. However, as the black hole grows, dark Bondi accretion eventually takes over in a second phase, rapidly consuming the entire SMFP core once $e^{t_{\rm acc}/\tau}\dot{M}_{\rm Bon}/\dot{M}_{\rm Edd}\big|_{M_{\rm GR}}\sim\mathcal{O}(1)$.

In Fig.~\ref{fig:acc_time}, we show the growth rate of the black hole mass due to dark Bondi accretion (blue), baryonic Eddington accretion (orange), and their combined contribution (red). Consider a halo of $M_{200} = 10^9\,\text{M}_\odot$ with concentration $c_{200} = 3$ at $z = 10$. Under the conservative estimate as discussed previously , the initial seed mass is $M_{\rm GR}=1{\rm\,\text{M}_\odot}$, $A/B^3=10^5$, and $\dot{M}_{\rm Bon}/\dot{M}_{\rm Edd}\big|_{M_{\rm GR}}f_{\rm Edd}\approx3.4\times10^{-5}$. As indicated by the solid curve, baryonic accretion dominates the growth until the black hole reaches $\sim10^4\, \text{M}_\odot$, assuming $f_{\rm Edd}=1$. After that point, dark Bondi accretion takes over, and the black hole quickly grows to $10^7\,\text{M}_\odot$ within $\sim500{\rm\,Myr}$, much faster than $\sim1500{\rm\,Gyr}$ it would require with dark accretion alone. We also show a case for more massive black holes, where we set $M_{\rm GR}=10^{3}{\rm\,M_\odot}$ under the same halo condition. As indicated by the dotted curve, the dark accretion completes within only $\sim200{\rm\,Myr}$, compared to $\sim1.5{\rm\,Gyr}$ without baryonic aid.

With dark accretion aided by baryons, a seed with a mass of $1\textup{--}10^3\,{\rm M_\odot}$ could grow into an SMBH within $200\textup{--}500\,{\rm Myr}$. In this scenario, the majority of the black hole mass originates from SIDM particles in the SMFP region of the halo rather than from baryonic matter; the predicted ratio of black hole to baryon masses may differ from that in scenarios involving only baryonic accretion. This accretion mechanism also applies to seeds originating from the remnants of Population III stars ($\sim100\,{\rm M_\odot}$)~\cite{Regan:2024wsu}, as indicated by the dashed curve.

We assume that accretion proceeds continuously without interruption. For light black hole seeds, a potential concern is that complex dynamics may cause them to wander, leading to inefficient accretion~\cite{Regan:2024wsu,Izquierdo-Villalba:2025aa}, unless they are embedded in dense environments such as star clusters~\cite{Alexander:2014}. In the scenario considered here, the initial seed is embedded in an extremely dense SMFP region of dark matter particles~\cite{Gu:2026zzq}, with a density exceeding that of a typical star cluster. In addition, during mergers, seed black holes from SIDM collapse or stellar remnants may sink toward the center due to dynamical friction in core-collapsed SIDM halos~\cite{Bosch:2026}. A more detailed study of dark accretion in such dynamical environments is deferred to future work.

{\bf Dark Eddington limit.} When a black hole becomes sufficiently massive, its dark Bondi accretion rate could become Eddington-limited due to dark radiation pressure, i.e., $\dot{M}_{\rm Bon}/\dot{M}_{\rm Edd} \lesssim 1$ in Eq.~\ref{eq:Bond_to_Edd}, where $\dot{M}_{\rm Edd}$ now refers to dark matter particles rather than baryons. Consider a generic SIDM particle model consisting of a fermionic dark matter particle $\chi$ that couples to a dark mediator $\phi$ with a coupling constant $g_\chi$ (e.g.,~\cite{Tulin:2017ara}). In this model, the self-scattering cross section $\chi\chi \rightarrow \chi\chi$ scales as $\sigma_{\chi\chi} \propto g^4_\chi m^2_\chi / m^4_\phi$, where $m_\chi$ is the dark matter particle’s mass and $m_\phi$ is the mediator’s mass. Meanwhile, the cross section for dark Thomson-like scattering off the mediator, $\chi \phi \rightarrow \chi \phi$, scales as $\sigma_{\chi\phi} \propto g_\chi^4 / m^2_\chi$ in the limit $m_\phi \ll m_\chi$. In this case, $\sigma_{\rm rad}/m$ in Eq.~\ref{eq:Bond_to_Edd} is replaced by $\sigma_{\chi\phi}/m_\chi$, corresponding to the dark radiation pressure.

For a viable SIDM model, the self-scattering cross section must be velocity-dependent across all halo mass scales from dwarfs to clusters, with an upper limit of $\sim0.1~\text{cm}^2/\text{g}$ inferred from cluster strong lensing ($M_{200}\sim10^{15}{\rm\,M}_\odot$) ~\cite{Kaplinghat:2015aga,Sagunski:2020spe,Andrade:2020lqq}. This indicates $m_{\phi}/m_\chi\lesssim10^{-3}$ for the model. For Bondi accretion, the temperature near the black hole could reach $\gtrsim 0.1 m_\chi c^2$~\cite{Shapiro:1983du}, so the dark mediator $\phi$ would be populated in the plasma. Although dark radiation pressure can be generated by the $\chi \phi \rightarrow\chi \phi$ scattering process, the associated opacity is extremely suppressed: $\sigma_{\chi\phi}/\sigma_{\chi\chi}\sim(m_\phi/m_\chi)^4\sim10^{-12}$.  For $\sigma_{\chi\chi}/m_\chi\sim10{\rm\,cm^2/g}$, $\sigma_{\chi\phi}/m_\chi\sim10^{-11}{\rm\,cm^2/g}$. Thus, from Eq.~\ref{eq:Bond_to_Edd}, we see that Bondi accretion of SIDM onto a black hole is not Eddington-limited for viable SIDM models. However, it could be relevant for SIDM models with dissipative interactions~\cite{Shen:2025evo}.

{\bf Discussion.} 
We discuss the mass function of LRDs predicted in the SIDM seeding scenario. The observed abundance of LRDs hosting black holes with masses $10^6\textup{--}10^8\,{\rm M}\odot$ at $z\sim5\textup{--}6$ is $\sim10^{-4}\textup{--}10^{-6}\,{\rm Mpc^{-3}\,dex^{-1}}$~\cite{Kokorev:2024kqm}. Using the Press--Schechter formalism~\cite{Press:1973iz,Jiang:2020rdj}, we estimate that the abundance of $10^9\textup{--}10^{10}\,{\rm M}\odot$ halos formed by $z\sim10$ is ${\rm d}n/{\rm d}\log_{10}M_{200}\sim1\textup{--}10^{-2}\,{\rm Mpc^{-3}\,dex^{-1}}$ in comoving volume. This halo abundance is substantially higher than the observed LRD abundance, as expected, since only a small subset of these halos can undergo core collapse rapidly enough to form black hole seeds at high redshifts. The collapse timescale depends sensitively on halo concentration, $t_{\rm cc}\propto c_{200}^{-7/2}$~\cite{Essig:2018pzq}, favoring early seeding in the high-concentration tail, while major mergers can further halt or delay core collapse. 

Accounting for these effects, Ref.~\cite{Jiang:2025jtr} finds that the resulting abundance of SIDM-seeded black holes is broadly consistent with the observed abundance of LRDs, based on a semi-analytical model of the dark matter-only case. The presence of baryons can accelerate the onset of core collapse and further shorten the collapse timescale, potentially increasing the fraction of halos that seed black holes by high redshifts. The magnitude of this effect depends on the baryonic distribution and assembly history, while the inner halo evolution may also be more complex than captured by the semi-analytical model. Cosmological hydrodynamical simulations incorporating SIDM, realistic halo assembly, and baryonic processes are therefore needed for a more robust prediction of the LRD abundance, which we leave for future work.

{\bf Conclusions.} We have investigated the formation of the JWST LRDs in the SIDM framework, in which the central dark matter halo undergoes gravothermal collapse to form a seed black hole. Even if the initially collapsing region that triggers relativistic instability has a mass on the order of one solar mass, the resulting seed can grow into an intermediate-mass black hole via Eddington accretion of baryonic gas, and subsequently evolve into an SMBH through dark Bondi accretion of SIDM particles. In this scenario, the majority of the black hole's mass originates from SIDM accretion rather than baryonic matter. In the future, dedicated numerical simulations are needed to test dark Bondi accretion and other SIDM accretion processes~\cite{Ostriker:1999ee,Hu:2005cd,Sabarish:2025hwb,Feng:2021qkj}. It will be important to accurately investigate the predicted population and clustering property of LRDs within this framework, e.g.,~\cite{Jiang:2025jtr,Roberts:2025poo,Roberts:2025yam,Wang:2026vzg,Zhang:2026uum} and explore the associated gravitational wave signatures~\cite{Shen:2025evo}. We may further improve the fluid simulations by incorporating anisotropic velocity dispersion~\cite{Fischer:2025rky,Gurian:2025zpc,Kamionkowski:2025fat}. 

The code used to model SIDM gravothermal evolution in the presence of a static baryonic potential is publicly available at~\href{https://github.com/ymzhong/gravothermal}{https://github.com/ymzhong/gravothermal}.

\textbf{Note added:} More recent simulations of SIDM accretion that account for non-equilibrium effects show that the accretion rate can exceed the Bondi rate at early times and subsequently become Bondi-like as the black hole grows and becomes dominant in the local gravitational potential~\cite{Gu:2026ode}; see also~\cite{Zeng:2026vyd}. This late-stage accretion is broadly consistent with our calculation. Thus, the overall timescale for dark accretion, without including baryonic accretion, estimated in our work, is likely conservative. Taken together, these results support the viability of the SIDM seeding scenario.

\bigskip
We thank Moritz Fischer, Fangzhou Jiang, Manoj Kaplinghat, Pranjal Ralegankar, Gowri Shanker, and Xuejian Shen for useful discussions. WXF was supported by Tsinghua's Shuimu Scholar Fellowship and the China Postdoctoral Science Foundation under Grant No.\,2024M761594. HBY was supported by the U.S. Department of Energy under Grant No.\,DE-SC0008541 and the John Templeton Foundation under Grant \# 63599. YZ was supported by the GRF grant\,11302824 from the Hong Kong Research Grants Council. The opinions expressed in this publication are those of the authors and do not necessarily reflect the views of the funding agencies.

\bibliographystyle{utphys}
\bibliography{acc}

\appendix
\clearpage
\onecolumngrid
\setcounter{page}{1}

\begin{center}
\textbf{\large --- Supplemental Material ---\\ $~$ \\
Dark Bondi Accretion Aided by Baryons and the Origin of JWST Little Red Dots}\\
\medskip
\text{Wei-Xiang Feng, Hai-Bo Yu, and Yi-Ming Zhong}
\end{center}
\setcounter{equation}{0}
\makeatletter
\renewcommand{\thesection}{S.\arabic{section}}
\renewcommand{\theequation}{S.\arabic{section}--\arabic{equation}}

\section{Gravitational Evolution of SIDM halos}
\label{sec:app1}

In our fluid simulations, we follow the procedure outlined in Refs.~\cite{Essig:2018pzq,Feng:2020kxv,Zhong:2023yzk} to study the gravothermal evolution of an SIDM halo under the influence of a baryonic potential. The simulation code is publicly available at~\href{https://github.com/ymzhong/gravothermal}{https://github.com/ymzhong/gravothermal}.

The evolution of the halo can be described by the following equations
\begin{eqnarray}
\frac{\partial M}{\partial r} = 4\pi r^2 \rho,\;
\frac{\partial (\rho \nu^2)}{\partial r} = - \frac{G (M+M_{B}) \rho}{r^2},\nonumber\\
\frac{\partial L}{\partial r} = -4 \pi \rho r^2 \nu^2 D_t \ln \frac{\nu^3}{\rho},\;
\frac{L}{4\pi r^2}=-\kappa\frac{\partial (m\nu^2)}{k_B\partial r},
\end{eqnarray}
where $M, \rho, \nu$, and $L$ are enclosed dark matter mass, density, 1D velocity dispersion, and luminosity profiles, respectively, and they are dynamical variables and evolve with time; $M_{B}(r)$ is the mass profile of baryons; $k_B$ is the Boltzmann constant, $G$ is the Newton constant, and $D_t$ denotes the Lagrangian time derivative. Heat conductivity of the dark matter fluid is given by $\kappa=(\kappa^{-1}_{\rm lmfp}+\kappa^{-1}_{\rm smfp})^{-1}$, where $\kappa_{\rm lmfp}\approx0.27C \rho\nu^3\sigma k_B/(Gm^2)$ and $\kappa_{\rm smfp}\approx2.1\nu k_B/\sigma$ denote conductivity in the long- and short-mean-free-path regimes, respectively, and $C\simeq0.75$ as calibrated from $N$-body simulations. In the short-mean-free-path regime, heat conduction can be characterized by the self-interaction mean free path $\lambda=1/\rho(\sigma/m)$ and $Kn=\lambda/H <1$, where $H=\nu/\sqrt{4\pi G\rho}$ is the scale height. In the long-mean-free-path regime, it is characterized by $H$ and $Kn >1$. 

Assuming an initial halo concentration $c_{200}$ with the given $M_{200}$ and cosmic redshift $z$, we translate to the scale density $\rho_s$ and radius $r_s$ in the NFW halo profile
\begin{equation}
\rho_{\rm NFW}(r)=\frac{\rho_s}{(r/r_s)\left(1+r/r_s\right)^2}\;.
\end{equation}
We then normalize all physical quantities using $\rho_s$, $r_s$, and their combinations, including the fiducial time $t_0 \equiv 1/\sqrt{4\pi G \rho_s}$, fiducial mass $M_0 \equiv 4\pi \rho_s r_s^3$, and fiducial velocity $\nu_0 \equiv r_s \sqrt{4\pi G \rho_s}$, for implementation in the simulation. 

We use a Plummer profile to model the baryon distribution in LRDs, 
\begin{equation}
\rho_B(r)=\frac{3M_B}{4\pi a^3}\left(1+\frac{r^2}{a^2}\right)^{-5/2}\frac{1+\cos \left( \frac{2\pi t}{P}\right)}{2},
\end{equation}
where $a$ and $M_B$ are the scale radius and mass of the profile, and $P$ is the oscillation period, respectively. We have checked that the Plummer profile provides a good approximation to the stellar distribution of the LRDs characterized by the S\'ersic profile. The parameters relevant to the baryons are $M_B/M_0$ and $a/r_s$ in the simulation. We set $P$ to be infinite for the static potential and $P=t_0$ for the oscillating potential.

We note that the density profile of dark matter halos, particularly at high redshifts, may deviate from the NFW profile. Our results are robust to such deviations, as dark matter self-interactions rapidly thermalize the inner halo, driving the system toward a quasi-equilibrium state (see, e.g.,~\cite{Tran:2025ycf}). We assume the presence of a baryonic potential but do not model its buildup history. This assumption is motivated by $N$-body simulation results, which show that for a given baryonic potential, the final SIDM halo evolution is largely insensitive to its assembly history~\cite{Zhong:2023yzk}.

\begin{figure*}[t]
\centering
   \includegraphics[width=0.44\textwidth]{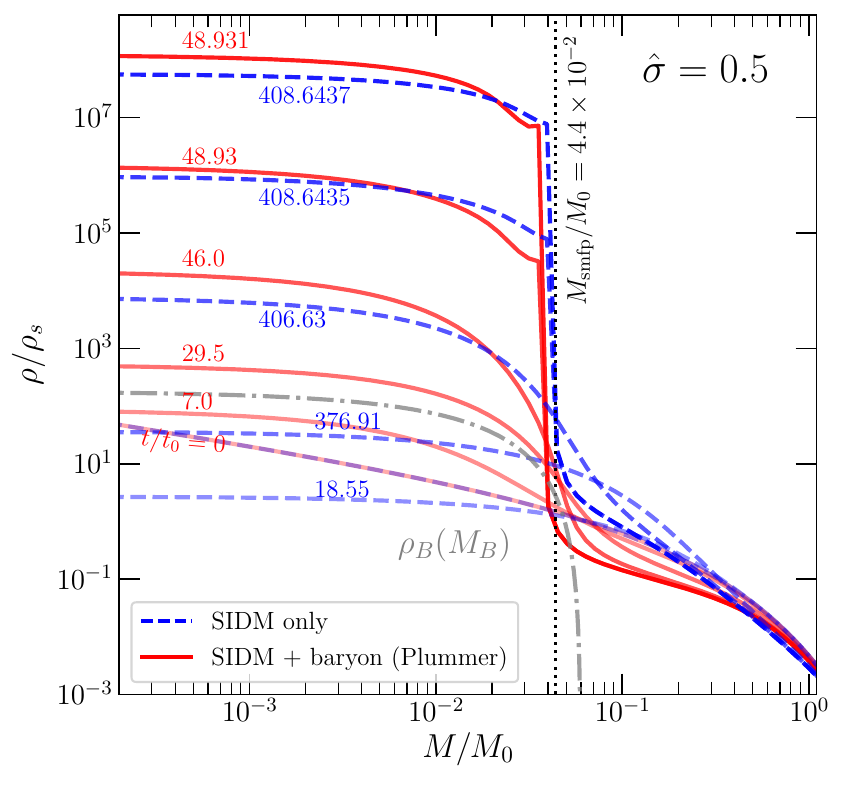}
   \includegraphics[width=0.44\textwidth]{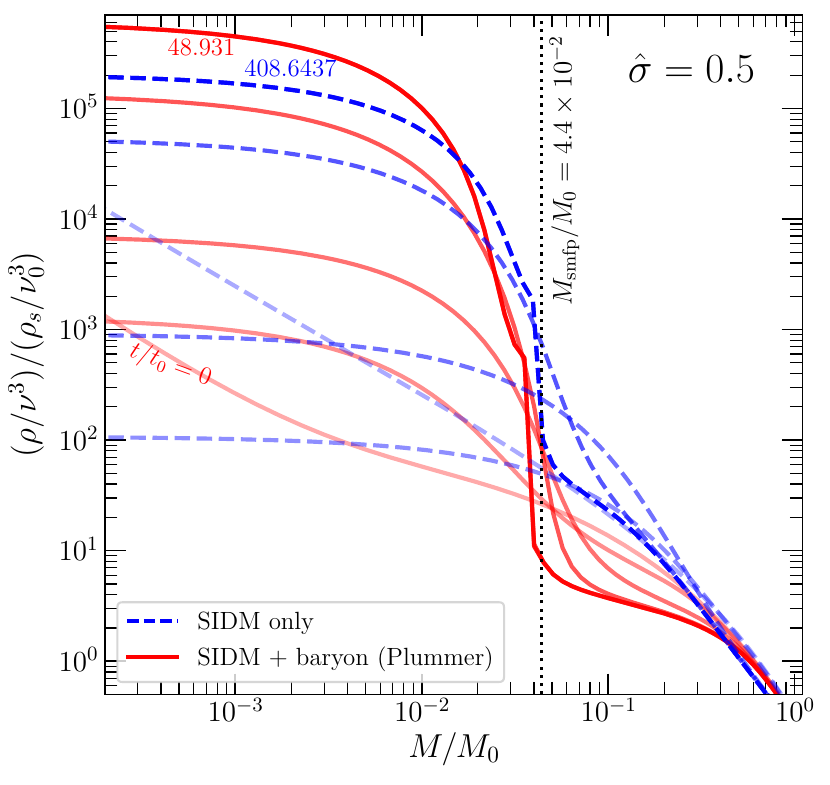}
   \includegraphics[width=0.44\textwidth]{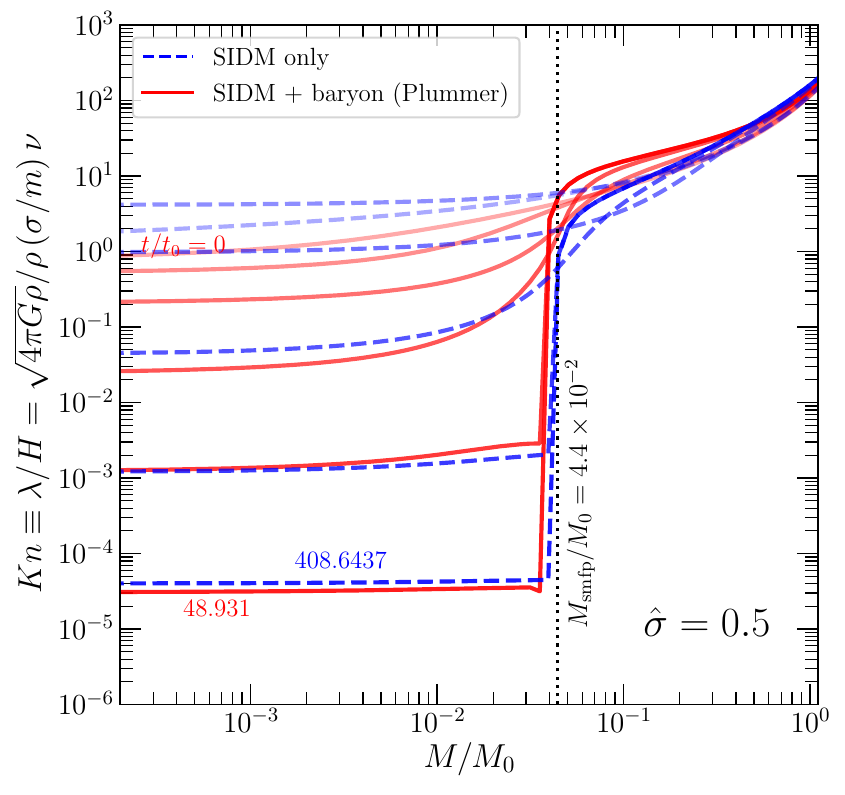}
   \includegraphics[width=0.44\textwidth]{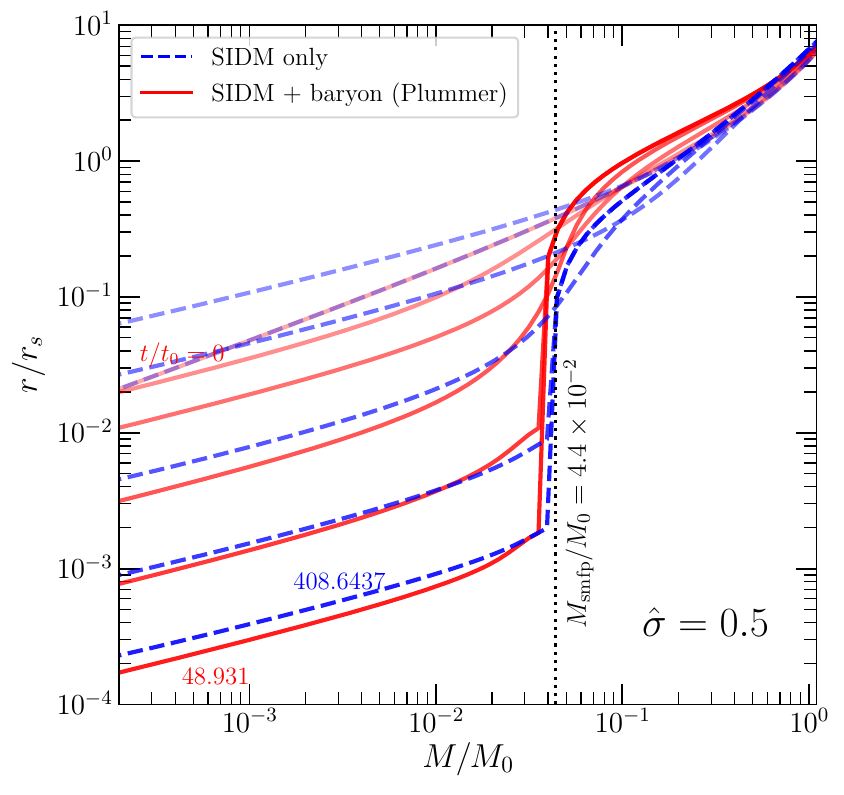}
    \caption{Evolution profiles for $\hat{\sigma} = \rho_s r_s (\sigma/m) = 0.5$. Shown are the density (top left), the ratio $\rho/\nu^3$ (top right), the Knudsen number (bottom left), and the radius (bottom right) as functions of enclosed mass at different stages of gravothermal collapse, with (solid red) and without (dashed blue) including the baryonic potential. The fiducial quantities are defined as $t_0 = 1/\sqrt{4\pi G \rho_s}$, $M_0 = 4\pi \rho_s r_s^3$, and $\nu_0 = r_s \sqrt{4\pi G \rho_s}$ in terms of the halo parameters $\rho_s$ and $r_s$. The dotted vertical lines mark the boundary of the short mean free path region, defined by $Kn = 1$. In the top left panel, the dash-dotted curve shows the baryon density profile. For the typical halo we consider for LRDs, $t_0\simeq10.9{\rm\,Myr}$ with $\rho_s\simeq1.5\times10^8{\rm\,M_\odot/kpc^3}$ and $r_s\simeq1.1{\rm\,kpc}$, yielding $\sigma/m=\hat{\sigma}\rho_s^{-1}r_s^{-1}\simeq15{\rm\,cm^2/g}$ for $\hat{\sigma}=0.5$. In the absence of a static baryonic potential, a short-mean-free-path core begins to develop at $t\gtrsim377t_0\simeq4.11{\rm\,Gyr}$ when $Kn\lesssim1$, and subsequently collapses with mass $M\simeq4.4\times10^{-2}M_0\sim10^7\textup{--}10^8{\rm\,M}_\odot$ at $t\simeq409t_0\simeq4.45{\rm\,Gyr}$. In contrast, in the presence of a baryonic potential, the inner region enters the short-mean-free-path regime much earlier, at $t\gtrsim7t_0\simeq76.3{\rm\,Myr}$, and ultimately collapses at $t\simeq49t_0\simeq530{\rm\,Myr}$.}
   \label{fig:profiles_s0d5}
\end{figure*}

\begin{figure*}[t]
\centering
   \includegraphics[width=0.44\textwidth]{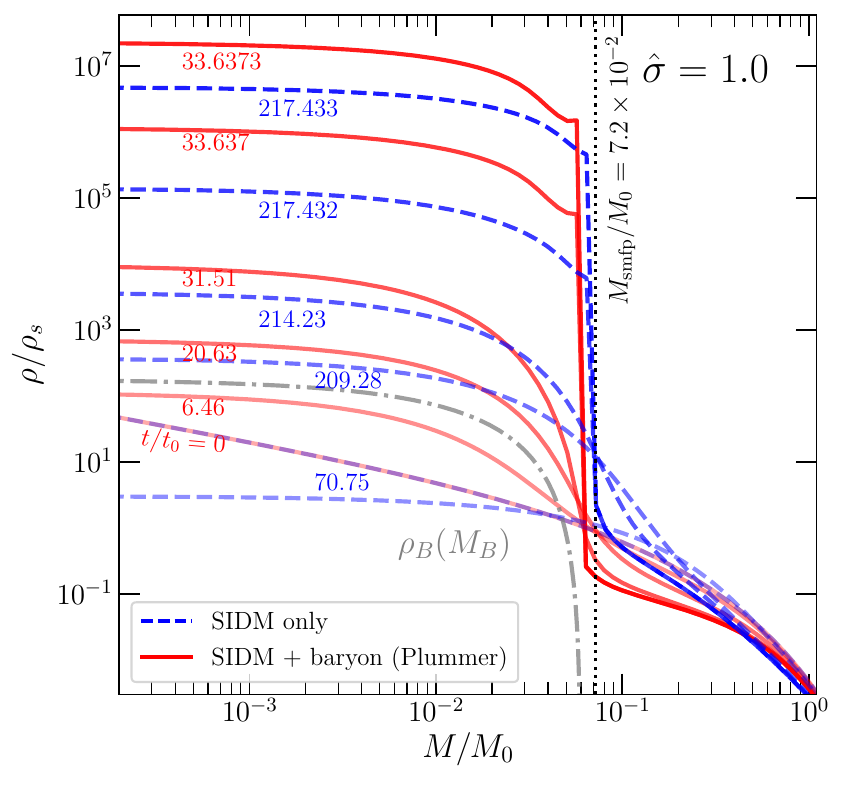}
   \includegraphics[width=0.44\textwidth]{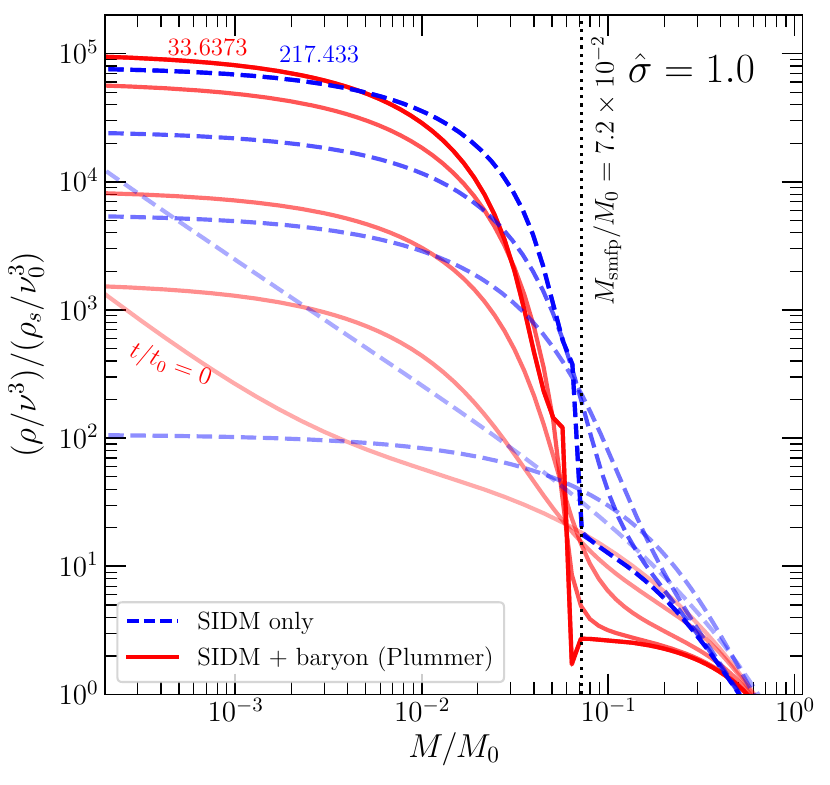}
   \includegraphics[width=0.44\textwidth]{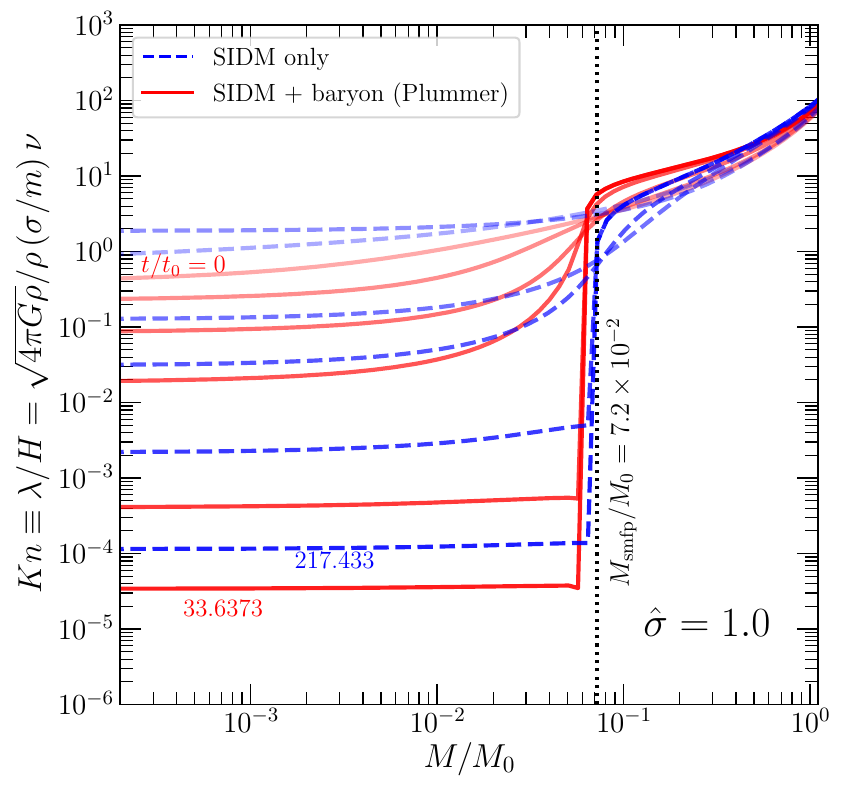}
   \includegraphics[width=0.44\textwidth]{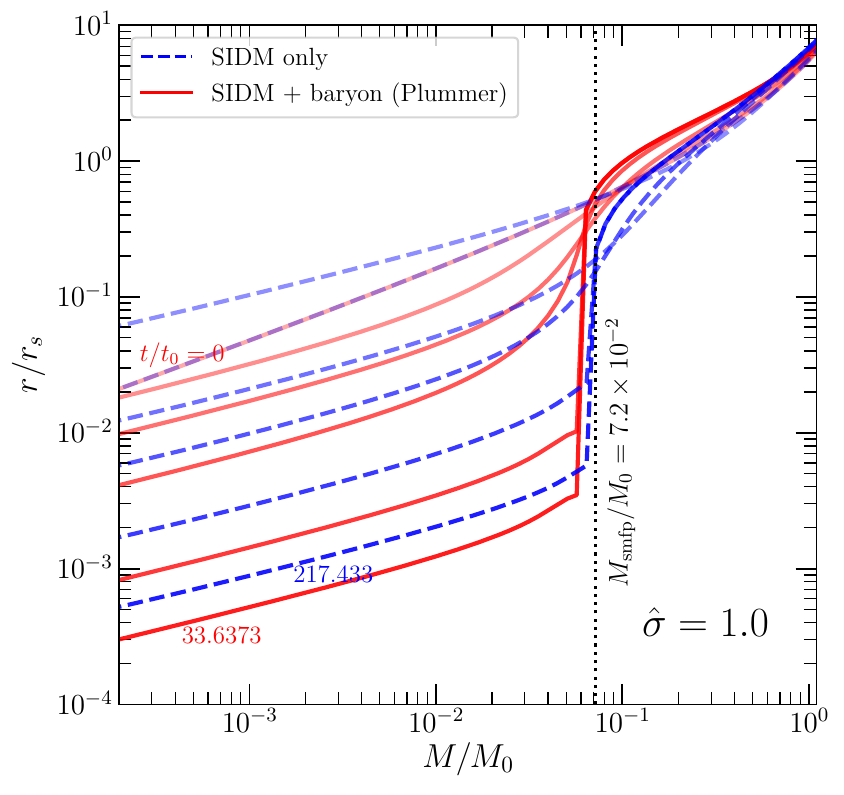}
    \caption{Similar to Fig.~\ref{fig:profiles_s0d5}, but for $\hat{\sigma} = (\sigma/m)\rho_s r_s = 1.0$, where $\sigma/m=\hat{\sigma}\rho_s^{-1}r_s^{-1}\simeq30{\rm\,cm^2/g}$ with $\rho_s\simeq1.5\times10^8{\rm\,M_\odot/kpc^3}$, $r_s\simeq1.1{\rm\,kpc}$, and $t_0\simeq10.9{\rm\,Myr}$. In the absence of a baryonic potential, a short-mean-free-path core begins to develop at $t\gtrsim209t_0\simeq2.28{\rm\,Gyr}$ when $Kn\lesssim1$, and subsequently collapses with mass $M\simeq7.2\times10^{-2}M_0\sim10^7\textup{--}10^8{\rm\,M}_\odot$ at $t\simeq217t_0\simeq2.37{\rm\,Gyr}$. In contrast, in the presence of a static baryonic potential, the innermost region is already in the short-mean-free-path regime even at $t/t_0$ close to $0$, and eventually collapses at $t\simeq34t_0\simeq370{\rm\,Myr}$.}
   \label{fig:profiles_s1d0}
\end{figure*}

Fig.~\ref{fig:profiles_s0d5} shows the density (top left), the ratio $\rho/\nu^3$ (top right), the Knudsen number (bottom left), and the radius (bottom right) as functions of enclosed mass at different stages of gravothermal collapse, with (solid red) and without (dashed blue) including the static baryonic potential. The dotted vertical lines mark the boundary of the short mean free path region, defined by $Kn = 1$. In the top left panel, the dash-dotted curve shows the baryon density profile. The simulation assumes $M_B/M_0 = 0.06$, $a/r_s = 0.1$, and $\hat{\sigma} = (\sigma/m)\rho_s r_s = 0.5$. Fig.~\ref{fig:profiles_s1d0} shows similar results, but for $\hat{\sigma} = 1.0$.

\end{document}